# Improved Calibration of Instruments for Small Direct Currents

Hansjörg Scherer[1], Dietmar Drung[2], and Christian Krause[1]

[1]Physikalisch-Technische Bundesanstalt (PTB), Bundesallee 100, 38116 Braunschweig, Germany
hansjoerg.scherer@ptb.de

[2]Physikalisch-Technische Bundesanstalt (PTB), Abbestraße 2-12, 10587 Berlin, Germany

**Abstract** — We report on new calibration methods for picoammeters and low-current sources. The 'Ultrastable Low-noise Current Amplifier' (ULCA) was used for the exemplary calibration of commercial state-of-the-art ammeter and current source instruments in the current range between 1 fA and 1 µA. The uncertainties achieved are compared with results obtained from two other calibration methods for small direct currents. It is shown that using the ULCA as calibrator allows achieving lower uncertainties together with the benefit of easier handling compared to alternative techniques. Also, the ULCA allows performing current meter and source calibrations between 1 fA and 1 µA with a single method.

## I. INTRODUCTION

Accurate measurements and calibrations in field of small electrical currents are of increasing importance due to needs from research (e.g. for high-accuracy single-electron transport experiments), environmental monitoring (e.g. for dust concentration measurements), medical and health metrology (e.g. in dosimetry), as well as from the industrial sector (e.g. in the microelectronics branch).
The ULCA has proven to be a powerful tool for the highly accurate traceable measurement and generation of small electrical currents [1] – [4]. Earlier, its performance as electrometer was evaluated and validated in a comparison with the 'capacitor charging' method used for current generation in the sub-nA range [5]. In this paper, the first exemplary applications of the ULCA for calibrations of low-current commercial state-of-the-art instrumentation, both for electrometers (picoammeters) as well as for low-current sources, are presented. The results show that the ULCA is excellently suited for this purpose, providing calibration and measurement capabilities excelling alternative state-of-the-art methods in terms of accuracy and practicability, and ease of operation.

## II. CALIBRATION PROCEDURES

A 'standard' version of the ULCA (single channel) was used for the calibrations. The instrument was equipped with a 3 GΩ/3 MΩ thin-film resistor network (first amplifier stage), providing 1000-fold current amplification, and a metal foil resistor (second stage) with $R_{IV}$ =1 MΩ, providing current-to-voltage conversion. It has a total effective transresistance of $A_{TR}$ = 1 GΩ (1000 × 1 MΩ) with an effective input current noise of 2.4 fA/√Hz [1].
Calibrations of device under test (DUT) instruments were performed under laboratory conditions. The ULCA temperature was monitored by its internal temperature sensor, and temperature coefficients of $A_{TR}$ and $R_{IV}$ were accounted in the evaluations. For currents between 1 fA and 5 nA, the ULCA was used in the current measurement and generation modes according to Fig. 3 in reference [1], using the full transresistance $A_{TR}$ = 1 GΩ. For metering and sourcing of higher currents between 5 nA and 1 µA, the ULCA was used in 'extended' current operation modes as presented in reference [6]. These modes only use the second amplifier stage with lower effective transresistance $R_{IV}$ = 1 MΩ.
Calibrations of $A_{TR}$ and $R_{IV}$ were performed with PTB's 14-bit cryogenic current comparator [2] before and during the calibration campaign. Due to the excellent stability of the ULCA, relative changes of both parameters were < 0.1 ppm over a period of 50 days. For the measurement evaluation, standard uncertainties for $A_{TR}$ and $R_{IV}$ were conservatively assigned 0.1 ppm including contributions from the calibration and short-term fluctuations.

*A. ULCA calibrating a picoammeter*

The ULCA configured in current source mode was used for calibrations of a commercial state-of-the-art 'sub-femtoamp source/meter' with its remote preamp being directly connected to the ULCA current output. The duration of the calibration measurements ranged between 1 h for currents ≥ 1 pA and 10 h for currents < 100 fA. The calibration current sourced by the ULCA was periodically reversed in order to suppress offset drifts.

*B. ULCA calibrating a current source*

The ULCA in electrometer mode was used for the calibration of a commercial state-of-the-art low-current 'calibrator/source', connected to the ULCA current input via a low-noise cable. The current sourced by the DUT was reversed once for each calibration in order to eliminate the current offset. The duration of the calibration measurements, limited by the onset of flicker noise caused by the DUT, were chosen 30 s for currents ≥ 10 nA, and 150 s for currents ≤ 1 nA.

## III. RESULTS

Fig. 1 shows the expanded uncertainties achieved for the DUT calibrations with the ULCA, in comparison with the uncertainties corresponding to the entries for the calibration and measurement capabilities (CMCs) of the Physikalisch-Technische Bundesanstalt (PTB) [7].

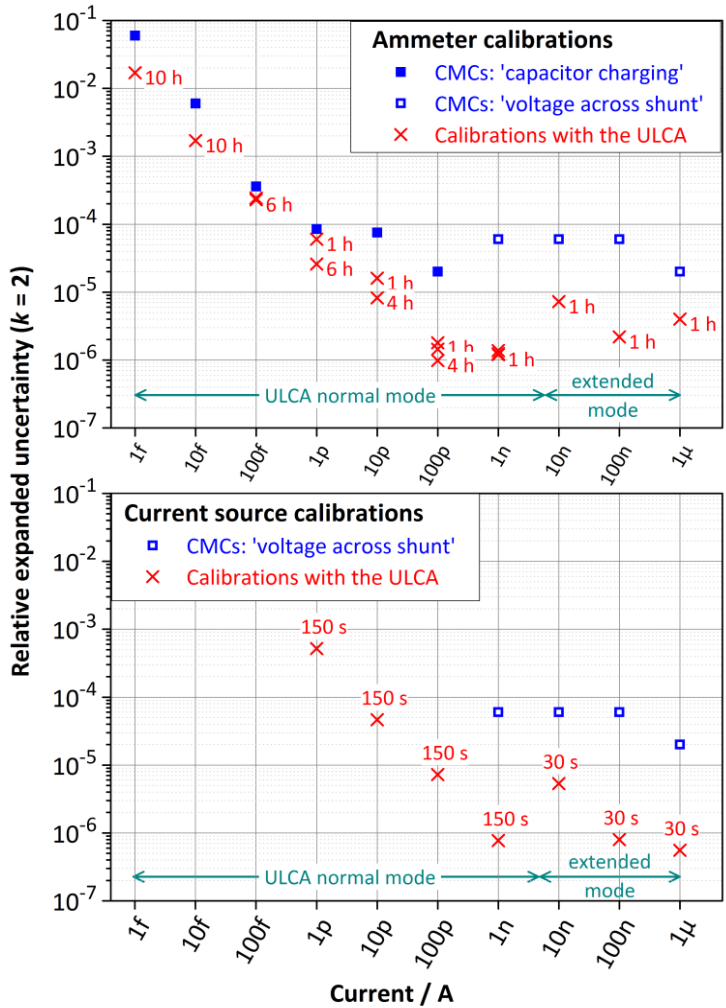

*Fig. 1. Total expanded uncertainties (coverage factor k = 2) for the calibrations of a commercial electrometer (upper panel) and a commercial current source (lower panel) by using the ULCA as calibrator current source and meter, respectively (cross symbols). Durations for each calibration measurement are denoted. Also, uncertainties corresponding to the actual CMC entries of PTB [7] are shown, achieved with the 'capacitor charging' (filled squares) and 'voltage across shunt' (open squares) methods.*

The results show that the ULCA enables calibrations of picoammeters and low-current sources at higher accuracy than alternative calibration methods, i.e. the 'capacitor charging' method and the 'voltage across shunt' method. In the current range from 100 pA to 1 µA, uncertainties for electrometer calibrations are decreased by one to two orders of magnitude at moderate measurement times. Current source calibrations can also be performed with significantly higher accuracy, and in an extended current range.

## IV. CONCLUSION

In summary, the ULCA provides several advantages compared to alternative established calibration methods for electrometers and low-current sources: i) calibration uncertainties can be reduced by up to two orders of magnitude, ii) only one calibrator instrument instead of two different setups is needed for calibrations over 9 decades of current, and iii) calibrations are simplified by the ease the ULCA offers in operation and handling.

Our results suggest the replacement of presently used calibration methods at PTB ('capacitor charging' and 'voltage across shunt' methods) by the ULCA in order to improve PTB's calibration and measurement capabilities significantly.

The excellent robustness of the ULCA provides temporal gain stability of typically 2 ppm/year [6], also maintained under harsh travelling conditions [3]. This makes it attractive as a travelling standard for inter-site comparisons as well as for applications in calibration service business. The ULCA will become available for customers after commercial launch in 2016.

## ACKNOWLEDGEMENTS

The authors thank M. Götz, E. Pesel, and U. Becker for calibrations of the ULCA, and B. Schumacher, C. Rohrig, and F. J. Ahlers for fruitful discussions.